\documentclass[prl,twocolumn,showpacs,preprintnumbers,amsmath,amssymb]{revtex4-1}
\makeatletter
\def\@dotsep{4.5}
\makeatother
\usepackage[dvips]{graphicx}
\usepackage{amsmath}
\usepackage{amsbsy}
\usepackage{color}
\usepackage{epstopdf}

\definecolor{textcolor}{cmyk}{0,0,0,1}
\definecolor{magenta}{rgb}{1,0,1}
\definecolor{green}{rgb}{0,1,0}
\definecolor{red}{rgb}{1,0,0}

\begin{document}

\title{ Gate-controlled conductance through bilayer graphene ribbons}
\author{J. W. Gonz\'alez$^1$, H. Santos$^2$,  E. Prada$^2$,
L. Brey$^2$ and L. Chico$^2$}
\affiliation{$^1$Departamento de F\'{i}sica, Universidad T\'{e}cnica Federico Santa Mar\'{\i}a, Casilla postal 110 V, Valpara\'{i}so, Chile
\\
$^2$Instituto de Ciencia de Materiales de Madrid, (CSIC),
Cantoblanco, 28049 Madrid, Spain}

\date{\today}
\begin{abstract}
We study the conductance of a biased bilayer graphene flake with monolayer nanoribbon contacts.
We find that the transmission through the bilayer ribbon strongly depends on the applied bias between the two layers and on the relative position of the monolayer contacts.
Besides the opening of an energy gap on the bilayer, the bias allows to tune the electronic density on the bilayer flake, making possible
the control of the electronic transmission by an external parameter.
\end{abstract}

\maketitle


The prospective use of graphene in nanoelectronics requires the possibility to
open gaps in its band structure in a controllable way. Due to the
chiral nature of the carriers \cite{Castro_Neto_RMP}, it is not
easy to open gaps and confine carriers in a single graphene
monolayer. Carbon-based structures as
nanoribbons \cite{Nakada_1996,Brey_2006b},
nanotubes \cite{Chico_1996b,Chico_1998,Santos_2009}, and graphene
bilayers \cite{Nilsson_2007,Castro_2007,Snyman_2007,Oostinga_2007,Fiori_2009a,Russo_2009}
are viable materials for nanoelectronics, since it is feasible to
change their electronic characteristics from semiconducting to metallic as a function of
geometric or external parameters.  Bilayer graphene is a
good candidate because a gap can be opened and controlled
 by an applied
bias
between its two layers.
Monolayer graphene nanoribbons (MGNs) stand out as optimal electrodes
for systems based on bilayer graphene,
with the aim of achieving the best integration of nanoelectronic components.
Thus, it is important to study the
electronic transport of bilayer graphene nanoribbons (BGNs) with MGN contacts.
Previous work has focused
on the electronic transport through bilayer graphene flakes in absence of
external gates \cite{Gonzalez_2010}. In such a case the conductance
shows strong oscillations as a
function of the energy of the incident electron and the length of
the bilayer region.
In this work we show that the conductance of BGNs connected to MGNs strongly depends on the
way the bilayer is contacted and on the applied gate voltage. This allows for an
external control of the electronic properties of the system.

\textit{Geometry.} We analyze electronic transport in the linear regime through a gated
BGN connected to two metallic MGN contacts. The monolayer leads can be either
 armchair or zigzag graphene nanoribbons
\cite{Nakada_1996,Brey_2006b}, serving as contacts to armchair or zigzag bilayer flakes respectively.
In both cases two configurations are possible: the bottom-bottom (1$\rightarrow$1) and
the bottom-top (1$\rightarrow 2$), where the ribbon leads are
connected to the same (1$\rightarrow$1) or to a different monolayer  (1$\rightarrow 2$) of the bilayer
flake. We consider BGN of width $W$ and length
$L$, and restrict our study to narrow nanoribbons in
the energy range for which only one incident electron channel is
active. The bias is applied symmetrically with respect to the top
($-V/2$) and the bottom ($V/2$) layers.

\begin{figure}[ht]
\includegraphics[width=\columnwidth,clip]{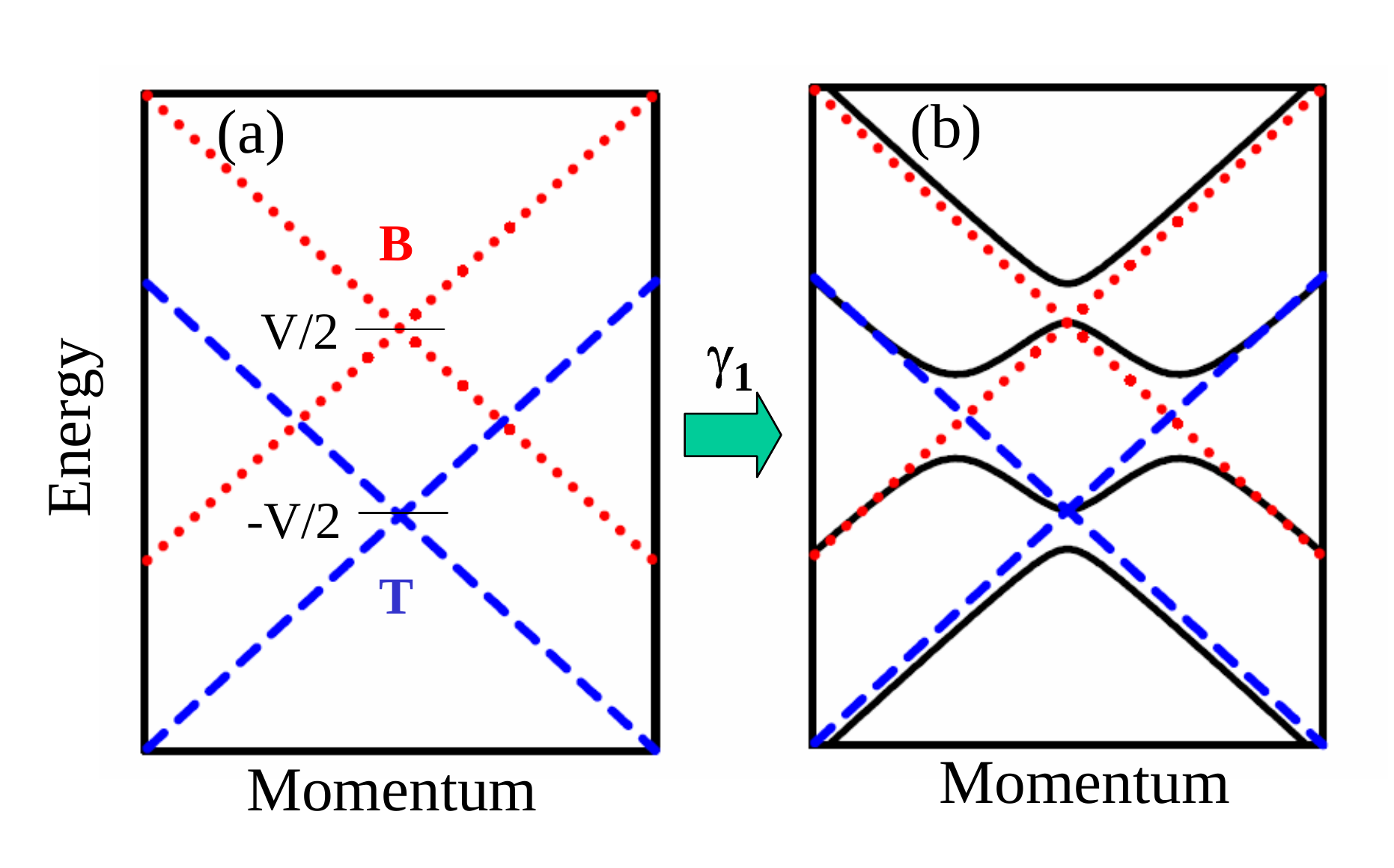}
\caption{(Color online) Schematic bandstructure of a biased graphene bilayer with and without interlayer coupling $\gamma_1$. (a) Biased bands without interlayer hopping. A positive (negative) bias $V/2$ is applied to the bottom (top) layer, producing a rigid shift of the corresponding linear dispersion relation (red dotted lines for the bottom layer, blue dashed lines for the top one). (b) When  $\gamma_1$ is switched on, gaps open at the band crossings, yielding the well-known Mexican hat shape and split-off bands of the bulk bilayer bandstructure. This picture allows to identify the top (T)/bottom (B) character of the different branches of the bilayer dispersion relation.}
\label{xxmex}
\end{figure}

\textit{Electronic structure of constituents.} The band structure of graphene has two inequivalent valleys. Within one valley, the low energy
properties of graphene are well described by the two dimensional Dirac equation,
$H=v_F \vec{\sigma} \cdot \vec{p} $, where $v_F \sim 1\times10
^6$ m/s is the Fermi velocity, $\vec{p}$ is the momentum operator relative to the Dirac point and
$\sigma _i$ are the Pauli matrices. The Dirac Hamiltonian acts on a
two-component spinor, $(\phi _A,\phi_ B)$, representing  the
amplitude of the wavefunction on the two inequivalent triangular
sublattices of graphene, labeled  $A$ and $B$. The band structure of armchair
graphene nanoribbons is obtained from the Dirac equation with the
appropriate boundary conditions \cite{Brey_2006b}. In all nanoribbons
the transverse momentum is quantized. For the armchair MGN case, when the number of
carbon atoms across the width of the ribbon is equal to $3m+2$, being $m$ a
positive integer, the smallest transverse momentum is zero. This yields the ribbon metallic with an energy dispersion $v_F p_x$, where $p_x$ is
the momentum along the nanoribbon.
%
%
\begin{figure}[ht]
\includegraphics[width=\columnwidth,clip]{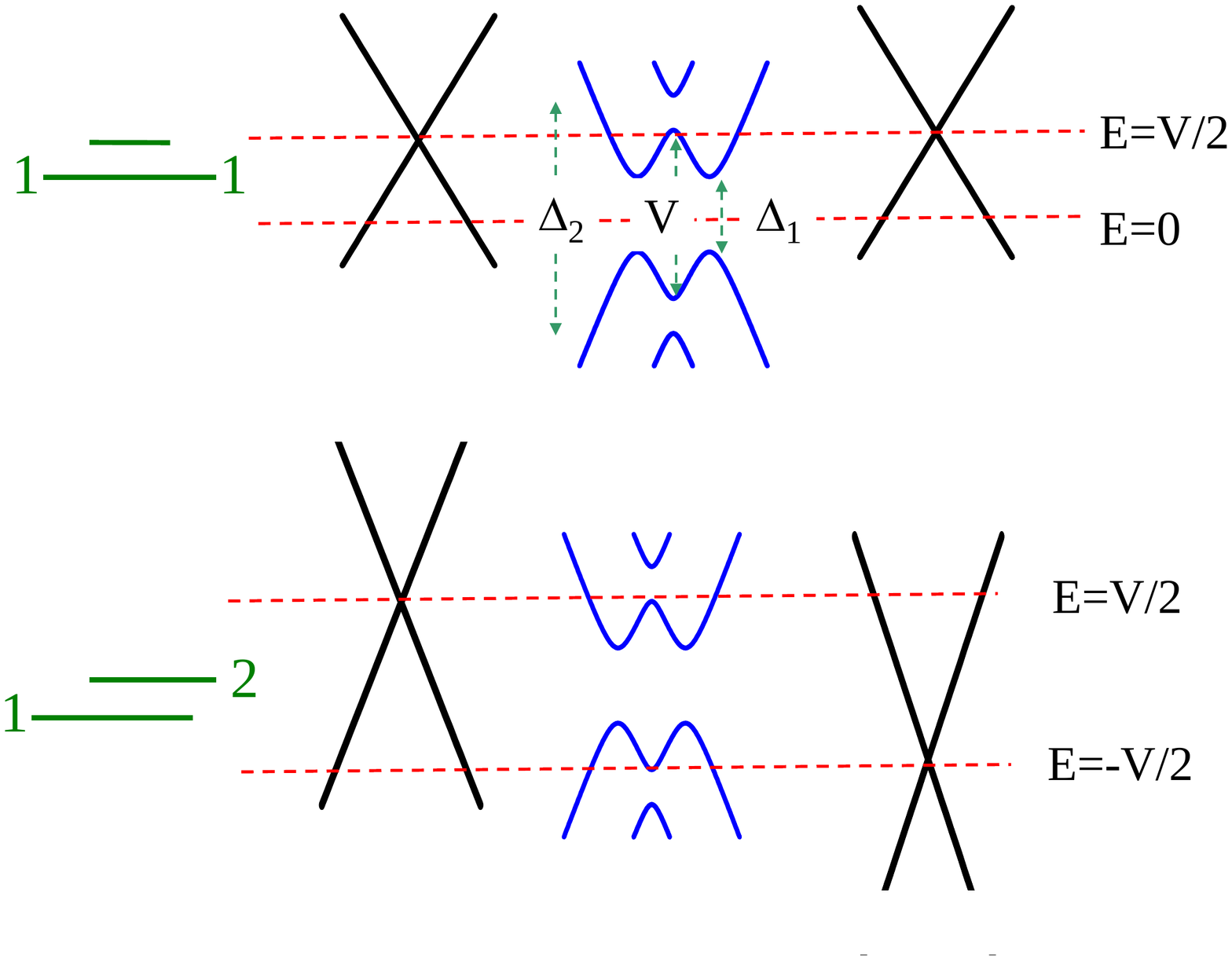}
\caption{(Color online) Schematic band structure of the bilayer armchair
nanoribbon with monolayer contacts.
Upper and lower panels
show the bottom-bottom (1$\rightarrow$1) and bottom-top
(1$\rightarrow$2) configurations.
 Left and right dispersions correspond to the
 metallic monolayer ribbons acting as electrodes, and the central dispersion
corresponds to the biased bilayer ribbon. The applied gate bias between
bottom and top layers is $V$.  For widths for which armchair
monolayer ribbons  are metallic, the band structure of the bilayer
is not affected by the confinement.  In the band structure of the
bilayer we indicate the three relevant gaps, $V$, $\Delta_1$, and
$\Delta _2$. } \label{SAC}
\end{figure}

%
 In addition to confined states,
zigzag MGNs support zero energy surface states located at the edges of the
ribbon \cite{Nakada_1996}.
%
%
In reciprocal space,
surface states occur between the two Dirac cones and their number
cannot be described by the two dimensional Dirac equation, which is only valid for the low
energy physics near the cones. Therefore, to describe zigzag graphene nanoribbons
we use a nearest-neighbor tight-binding Hamiltonian
$H=-t\sum ( a^+_i b_j +h.c.)$. Here $a_i (b_i)$ annihilates an
electron on site $i$ of sublattice $A(B)$, and the hopping parameter
$t$ is related to the Fermi velocity by $v_F=\frac
{\sqrt{3}} 2 a t$, where $a$ is the graphene lattice constant,
$a\approxeq 2.46$ \AA .

\begin{figure}[ht]
\includegraphics[width=\columnwidth,clip]{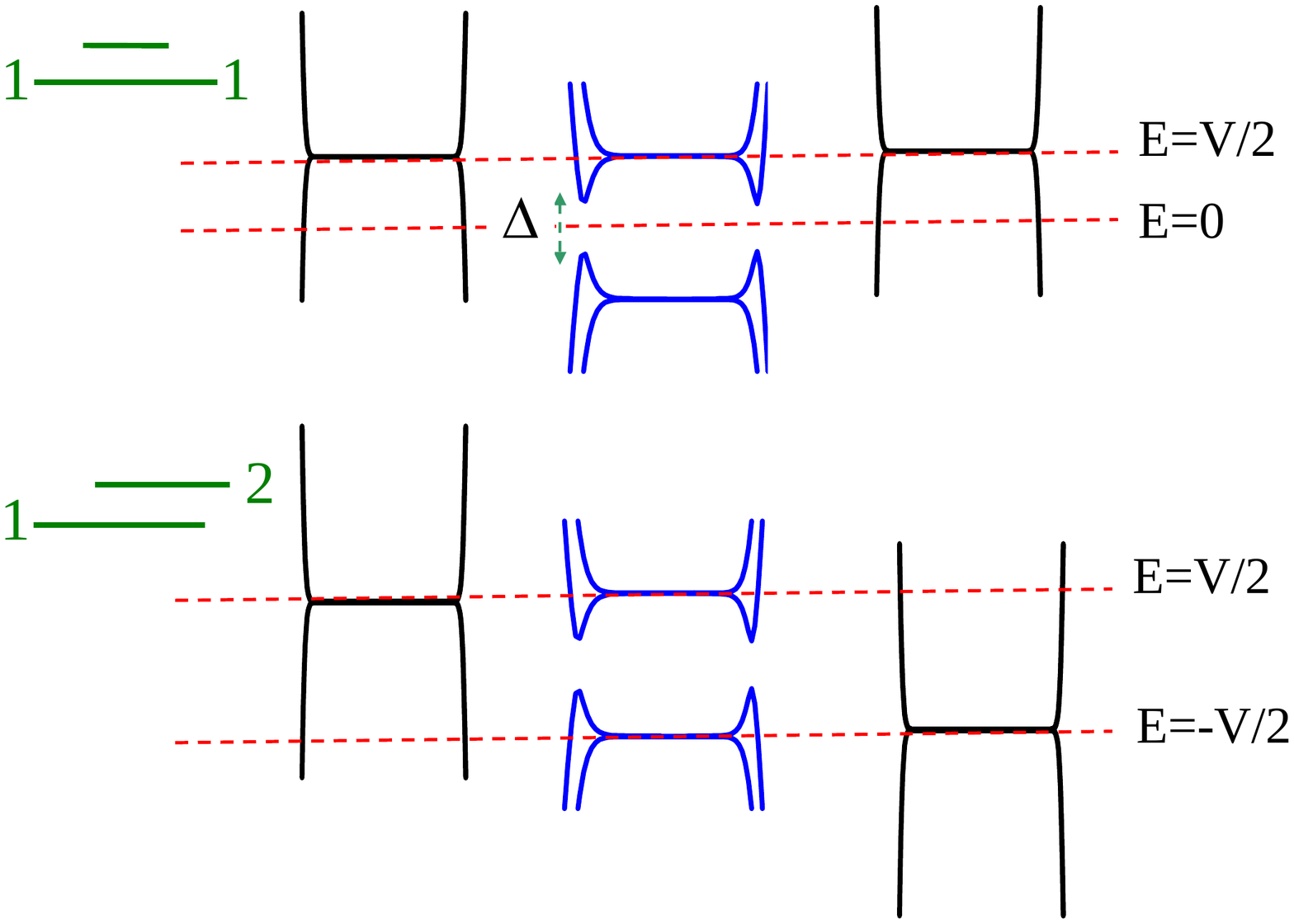}
\caption{(Color online) Same as Fig.\ref{SAC} but for zigzag
terminated nanoribbons.  For zigzag monolayer nanoribbons, states in
different valleys transport charge in opposite directions. In the
biased bilayer ribbon, central part, edge localized states with
energies $\pm V/2$ appear connecting the Dirac points. At the Dirac
points of the bilayer ribbons, topological surface states exist in
the gap. In wide ribbons these surface states close the energy gap.
In narrow bilayer ribbons states in different edges interact and
open a gap $\Delta$ in the spectrum. } \label{SZZ}
\end{figure}

Bilayer graphene consists of two coupled graphene layers with
inequivalent sites $A_1, B_1$ and $A_2 , B_2$ on the bottom and top
layers respectively. We consider the Bernal stacking in which the
$B_2$ sublattice is exactly on top of the sublattice $A_1$. Within one valley, the low
energy properties of a biased bilayer graphene are well
described \cite{McCann_2006} by the Hamiltonian
\begin{equation}
H_{BG}=v_F \tau_0 \otimes \vec{\sigma }\cdot \vec{p} + \frac
{V} 2 \tau _z \otimes \sigma_0 + \frac {\gamma_1}2 ( \tau _x
\otimes \sigma _x - \tau _y \otimes \sigma _y) \label{hamil_BL},
\end{equation}
where $\gamma_1 \sim t/10 $ is the hopping parameter between the
closest carbon atoms belonging to different layers, $\sigma _i$ are again the Pauli matrices for the sublattice degree of freedom and $\tau _i$ are
the Pauli matrices for the layer index ($\sigma_0$ and $\tau_0$ are the unit matrices in both subspaces).  This
Hamiltonian acts on the four-component spinor $(\phi
^1 _A,\phi ^1_ B, \phi
^2 _A,\phi ^2_ B)$ representing  the amplitude of the wavefunction on the two
sublattices  $A$ and $B$ of the two layers $1$ and $2$. The energy
bands are
$ \epsilon_{BG} ^2 ( p ) = v_F^2 p^2 + \frac {V  ^2} 4 + \frac
{\gamma_1^2} 2 \pm \frac 1 2 \sqrt{4 v_F^2 p^2 ( V   ^2+ \gamma_1
^2)+ \gamma_ 1 ^4} $.
The low energy  band has a Mexican hat shape with a minimum gap
$\Delta _1 = \gamma _1  |V|/(\sqrt {\gamma _1 ^2 + V^2})$, see Fig. \ref{xxmex}.
The minimum gap of the second subband
is $\Delta _2 = 2 \sqrt {\gamma _1 ^2 + ( \frac  V  2 ) ^2}$ and occurs at  $p=0$.
Fig. \ref{xxmex} illustrates how the Mexican hat shape and the split-off bands arise:
without interlayer hopping, the applied bias shifts the linear band dispersions of the two layers;
the interaction between layers opens gaps at the intersections of the bands.

The electronic structure of an
 armchair BGN depends
on the width of the ribbon. As in the monolayer case, when the
number of carbon atoms along a BGN layer is equal to $3m+2$,
the smallest transverse momentum is zero, and the dispersion of the
armchair BGN is $\epsilon_{BG}  ( p _x )$,  $p_x$ being the momentum along
the ribbon, see Fig. \ref{SAC}. In the case of zigzag biased
BGNs the system supports two kinds of surface states \cite{Castro_2008}:
(i) states with energies $ \sim \pm V/2$, similar to those
occurring in  zigzag MGNs,  and (ii) valley-polarized states with
energies in the gap. At each edge of the ribbon there are two
surface states carrying current in opposite directions and belonging
to different valleys. These states have a topological
nature \cite{Martin_2008} but the metallicity of the
edge is not protected against intervalley
scattering nor against
interedge intravalley scattering. The latter occurs when the ribbon width
is smaller than the penetration length of the surface
states \cite{Morpurgo}, $\ell \sim \sqrt{3} \frac t {\gamma_1} a$,
that for realistic values of the interlayer hopping is around 17$a$. This large
$\ell$ value produces
an interedge
scattering gap $\Delta$ in the spectrum of narrow zigzag BGNs, see
Fig. \ref{SZZ}. Although the valley-polarized surface states can be
modeled with the Dirac Hamiltonian, the coupling between states
localized in opposite edges is better described using a
tight-binding Hamiltonian, which takes into account the coupling between inequivalent
Dirac points.

\textit{Electronic conductance.} \textit{Armchair nanoribbons.} As
discussed above, the Dirac Hamiltonian describes appropriately the
low energy band structure of armchair nanoribbons. Therefore, we
calculate the conductance of the system by matching the
eigenfunctions of the Dirac-like Hamiltonian.  Given an incident
electron coming from the left monolayer ribbon and with energy $E$,
we compute the transmission coefficient to the right monolayer lead.
The boundary conditions on the wavefunctions determine the value of
the transmission. In the bottom-bottom configuration (1$\rightarrow$1) the
wavefunctions of the bottom layer $\phi_A ^1$ and $ \phi_B ^1$ should be continuous
at the beginning ($x=0$) and at the end ($x=L$) of
the
bilayer
region. For the top layer the wavefunction
should vanish in one sublattice at $x=0$ and on the other
sublattice at $x=L$, e.g.,  $\phi_A ^2 (x=0)=\phi_B ^2 (x=L)=0$. In the
bottom-top configuration the bottom wavefunctions $\phi_ \mu ^1$
and the top wavefunctions $\phi_ \mu ^2$ should be continuous at
$x=0$ and $x=L$ respectively. In addition, the
hard-wall condition should be satisfied, $\phi_B ^2 (x=0)= \phi_A ^1 (x=L)=0$.
\textit{Zigzag nanoribbons.} To describe adequately the low
energy properties of  zigzag nanoribbons in the full Brillouin
zone it is necessary to use a tight-binding Hamiltonian.
We use a  Green's
function approach to obtain the transport properties \cite{Chico_1996b,Datta_book,Gonzalez_2009}.
 In this method the system is
divided in three parts, namely, a finite-size bilayer section connected to the
right and left monolayer semi-infinite leads. The Green's function of the central
region is
\begin{equation}
\mathcal{G} _C (E) = ( E- H _C - \Sigma _ L - \Sigma _ R ) ^{-1} \, \,\, \, ,
\end{equation}
where $H_C$ is the bilayer Hamiltonian and $\Sigma _L$ and $\Sigma _R$ are the selfenergies at the ends of the bilayer region due to the presence of the leads. The selfenergies contain the information on the type of connection, i.e., 1$\rightarrow$1 or 1$\rightarrow$2, of the system. In the linear regime, the conductance is given by
\begin{equation}
G= 2 \frac {e ^2} h T(E) = 2 \frac {e ^2} h Tr [ \Gamma _L  \mathcal{G}_C \Gamma _R \mathcal{G} ^+ _C ] \, \, \, ,
\end{equation}
where $T(E)$ is the transmission at the Fermi energy $E$ and $\Gamma _L$ and $\Gamma _R$ are the couplings between the bilayer and the left and right monolayer leads respectively.

\begin{figure}[ht]
\includegraphics[width=\columnwidth,clip]{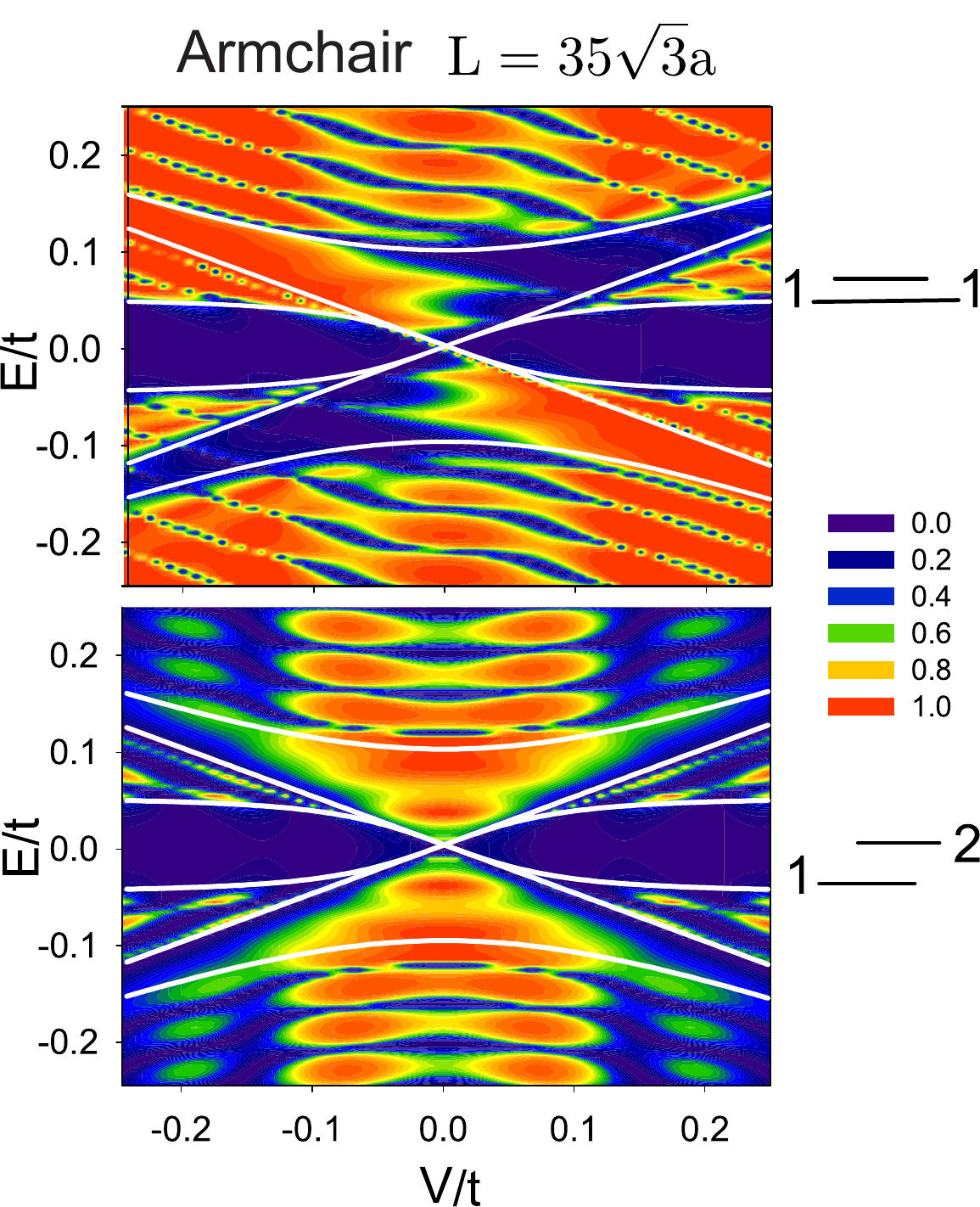}
\caption{(Color online) Transmission as a function of energy and
applied voltage for armchair nanoribbons in the 1$\rightarrow$1
and 1$\rightarrow$2
configuration. The value of the interlayer hopping
parameter is $\gamma _1 =0.1t$. The length of the
bilayer
region
is $L=35\sqrt{3}a$. The values of
the gap edges $\pm \Delta _1 /2$, $\pm \Delta _2/2$ and $\pm V/2$, are plotted with white lines,
see Fig.\ref{SAC} . } \label{res_AC}
\end{figure}

\textit{Results.} \textit{Armchair nanoribbons.} In
Fig. \ref{res_AC} we plot the transmission as a function
of the incident energy and the applied gate voltage for an armchair nanoribbon
system in the 1$\rightarrow$1
and 1$\rightarrow$2
configurations. The length of the bilayer flake is $L=35\sqrt{3}a$
and  the value of the interlayer hopping is $\gamma _1 =t/10$. The
results are independent of the width of the ribbon, provided that the
monolayer ribbons are metallic and that the energy of the incident
electron is lower that the energy of the second subband. The
transmission is obtained in the continuum approximation, but we have
checked that the results coincide with those obtained within the tight-binding
approach. Due to the symmetry of the contacts, the 1$\rightarrow$2
configuration shows electron-hole and $V \rightarrow -V$  symmetry.
This is not the case for the
1$\rightarrow$1
 configuration, for which
the location of the contacts precludes those symmetries. In both
cases the conductance is suppressed for energies in the gap
$E < |\Delta _1 |/2$,  for which there are no available
states for the conductance in the bilayer region. We first discuss the results for
the bottom-bottom configuration. 
In the energy window $\Delta _1 /2 < |E| <
|V|/2$ there are two propagating states
in the bilayer part and the
conductance is finite. In the range of energies $|V|/2 < |E| <\Delta_2 /2 $ there is only one propagating mode in the central region;
 but
in this configuration,
when $E$ and $V$ have the same sign, this mode is mostly located in the opposite (top)
layer to the leads (bottom), as can be seen in Fig. \ref{xxmex}(b), and the conductance is near zero.
Thus, by applying a gate voltage between the two layers we can tune the electronic density in the bilayer.
Changing the distribution of carriers from one layer to the other allows to control the conductance of the
system by means of an external parameter.
 For
energies $|E|>\Delta _2 /2$ the transmission is finite with
antiresonances associated with interferences in the bilayer region
due to the existence of two propagating channels. \cite{Gonzalez_2010}
These interferences are weaker for voltages $|V| >  \gamma _1$, with and overall
nonzero conductance, because in this case the incident current from the left electrode
is transmitted efficiently to the upper branch of the bilayer dispersion relation, with bottom character, and from there to the right (bottom)
lead, with an almost perfect wavevector matching \cite{Chico_2004}.
The weak interferences are due to the
bilayer-confined states arising from the coupling to the top layer flake.
Note
the linear dependence of the position of the antiresonances on the
applied voltage: the energy of the confined states in the top layer are displaced by the applied bias $-V/2$,
thus changing the occurence of the antiresonances correspondingly.

 In the bottom-top configuration the conductance is
not suppressed for $|V|/2 < |E| <\Delta_2 /2$ because in this
case the incoming and outgoing electrons belong to
different layers: the propagating mode in the bilayer has a predominant top character (see Fig. \ref{xxmex}), being easily transmitted to
the right electrode.
For this configuration, the transmission at energies $|E|>\Delta _2 /2$ is generally suppressed, even though there
are two propagating modes in the bilayer. This can be understood by noticing the wavevector mismatch \cite{Chico_2004} between left and right electrodes
produced by the applied bias, as depicted in Fig. \ref{xxmex}.
Away from the gap
 the transmissions
in the
1$\rightarrow$1
and
1$\rightarrow$2
configurations are rather
complementary; the antiresonances that occur in the
1$\rightarrow$1
configuration become resonances in the
1$\rightarrow$2
case. This
complementarity of the conductance  can be understood by resorting
to a simple non-chiral model. Consider an incident carrier, with
energy larger than the gap, coming from the left and therefore in the bottom sheet. When
arriving at the bilayer central region, the incident wavefunction
decomposes into a combination of the two eigenstates of the biased
bilayer. The conductance through the bilayer region is proportional
to the probability of finding an electron at the top (bottom) end of
the central region for the bottom-top (bottom-bottom) configuration.
As the total probability of finding the electron
at the end of the bilayer region is unity, the bottom-bottom and the
bottom-top transmissions should be the opposite.

\begin{figure}[ht]
\includegraphics[width=\columnwidth,clip]{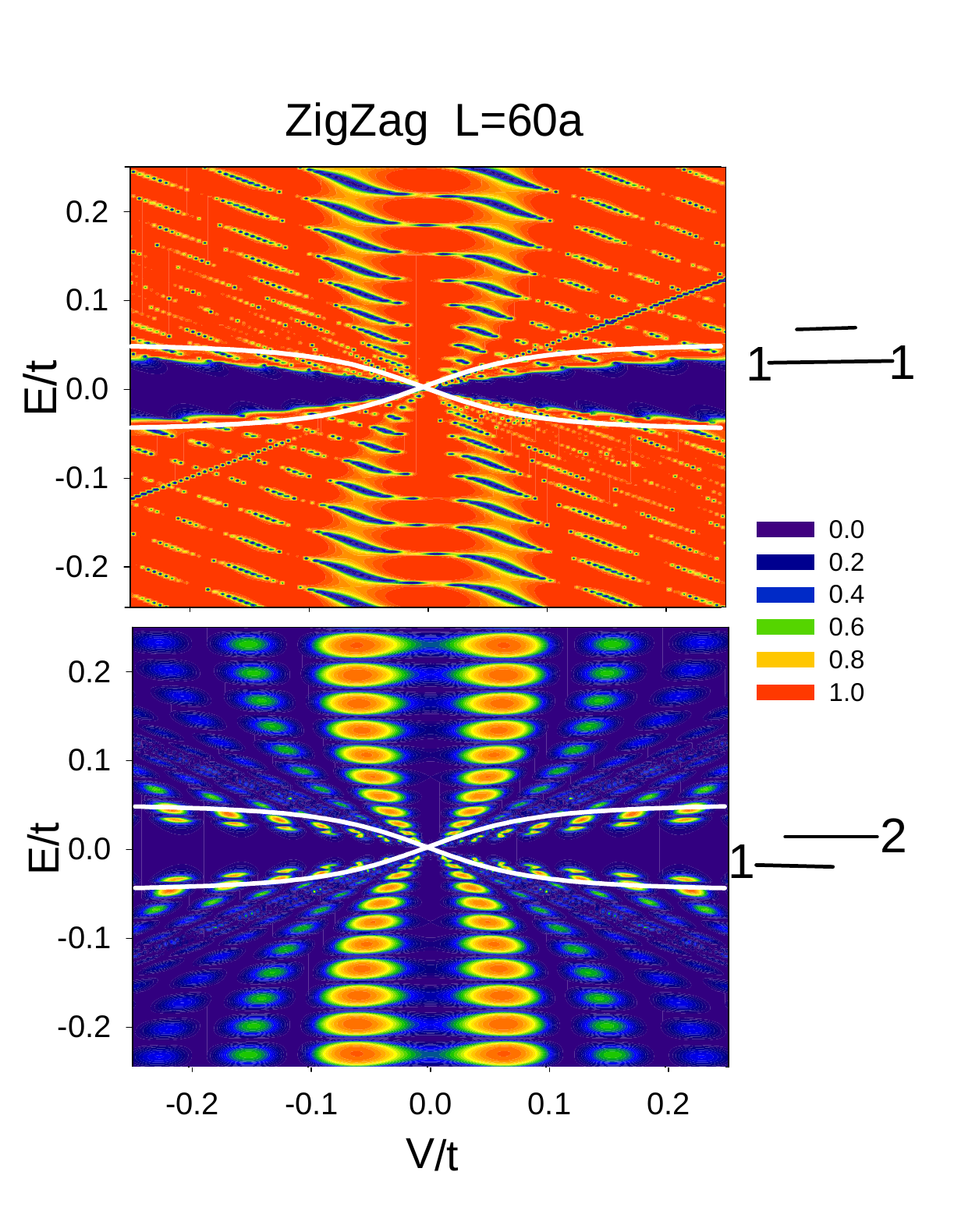}
\caption{(Color online) Transmission as a function of energy and
applied voltage for zigzag GNRs
in the 1$\rightarrow$1
and 1$\rightarrow$2
configurations. The interlayer hopping
parameter is $\gamma _1 =0.1t$. The number of atoms across the
nanoribbon is 16, corresponding to a width $W=11a/\sqrt{3}$. The
bilayer region length
 is $L=60a$.
  The values of
the gap edges $\pm \Delta _1 /2$
are plotted in white lines,
see Fig. \ref{SAC}.}
\label{res_ZZ}
\end{figure}
\par\noindent
\textit{Zigzag nanoribbons.} In Fig.\ref{res_ZZ} we plot the
transmission as function of the Fermi energy and
the applied gate voltage for zigzag nanoribbons in
the
1$\rightarrow$1
and
1$\rightarrow$2
configurations. These results have
been obtained using a tight binding Hamiltonian and a recursive Green's
function technique \cite{Chico_1996b,Datta_book,Gonzalez_2009}.
The conductance of zigzag graphene nanoribbons depends on the ribbon width.
The results presented in Fig.\ref{res_ZZ} correspond to a narrow ribbon,
$W=11a/ \sqrt{3}$, for which there is only one channel coming from the left contact for all the plotted energies. As for the armchair-based systems, there is a
 strong  complementarity between the
1$\rightarrow$1
and the
1$\rightarrow$2
configurations, yielding a
very different conductance as a function of the energy and bias for the two configurations.
Other features, as antiresonances (resonances) in the bottom-bottom
(bottom-top) configuration are similar to those occurring  in armchair
nanoribbons and have the same origin.
In the upper panel of  Fig. \ref{res_ZZ}, corresponding to the  1$\rightarrow$1  configuration, the previous gapped region in the armchair case between $V/2$ and $\Delta_2/2$ has shrunk to a line of slope $V/2$. This is easy to understand by observing the zigzag bandstructure of Fig. \ref{SZZ}. Another remarkable feature in Fig. \ref{res_ZZ} is
the existence of a transport gap $\Delta$ smaller than the bulk
gap $\Delta _1$ of the gated bilayer. As mentioned above, the gap
$\Delta$ appears because of the coupling between
states with the same valley polarization
localized in different edges and moving in
opposite directions. The penetration length $\ell$ of these
surface states is rather large; for nanoribbons narrower than
$\ell$ this produces a noticeable transport gap
smaller than the bulk gap.

In summary, we have calculated the conductance of bilayer graphene flakes with monolayer
nanoribbon contacts with a bias voltage between layers. Depending on the position of the electrodes and
on the applied bias, there is a strong variation of the conductance. Besides the energy gap
opened by the bias, the conductance can be tuned by changing the spatial distribution of the carriers in the bilayer region, thus
allowing for the external control of the transport through graphene bilayer flakes.

%
\textit{Acknowledgments.}
This work has been partially supported by MEC-Spain under grant
FIS2009-08744. J.W.G.
would like to gratefully acknowledge helpful discussions with Dr. L.
Rosales, the ICMM-CSIC for their hospitality and the financial support of MESEUP research
internship program, CONICYT (CENAVA, grant ACT27) and USM 110856 internal grant.


\begin{thebibliography}{20}
\expandafter\ifx\csname
natexlab\endcsname\relax\def\natexlab#1{#1}\fi
\expandafter\ifx\csname bibnamefont\endcsname\relax
  \def\bibnamefont#1{#1}\fi
\expandafter\ifx\csname bibfnamefont\endcsname\relax
  \def\bibfnamefont#1{#1}\fi
\expandafter\ifx\csname citenamefont\endcsname\relax
  \def\citenamefont#1{#1}\fi
\expandafter\ifx\csname url\endcsname\relax
  \def\url#1{\texttt{#1}}\fi
\expandafter\ifx\csname urlprefix\endcsname\relax\def\urlprefix{URL
}\fi \providecommand{\bibinfo}[2]{#2}
\providecommand{\eprint}[2][]{\url{#2}}

\bibitem[{\citenamefont{Castro-Neto et~al.}(2009)\citenamefont{Castro-Neto,
  F.Guinea, N.M.R.Peres, K.S.Novoselov, and A.K.Geim}}]{Castro_Neto_RMP}
\bibinfo{author}{\bibfnamefont{A.~H.} \bibnamefont{Castro-Neto}},
  \bibinfo{author}{\bibnamefont{F.Guinea}},
  \bibinfo{author}{\bibnamefont{N.M.R.Peres}},
  \bibinfo{author}{\bibnamefont{K.S.Novoselov}}, \bibnamefont{and}
  \bibinfo{author}{\bibnamefont{A.K.Geim}}, \bibinfo{journal}{Rev.\ Mod.\
  Phys.} \textbf{\bibinfo{volume}{81}}, \bibinfo{pages}{109}
  (\bibinfo{year}{2009}).

\bibitem[{\citenamefont{Nakada et~al.}(1996)\citenamefont{Nakada, Fujita,
  Dresselhaus, and Dresselhaus}}]{Nakada_1996}
\bibinfo{author}{\bibfnamefont{K.}~\bibnamefont{Nakada}},
  \bibinfo{author}{\bibfnamefont{M.}~\bibnamefont{Fujita}},
  \bibinfo{author}{\bibfnamefont{G.}~\bibnamefont{Dresselhaus}},
  \bibnamefont{and} \bibinfo{author}{\bibfnamefont{M.~S.}
  \bibnamefont{Dresselhaus}}, \bibinfo{journal}{Phys. Rev. B}
  \textbf{\bibinfo{volume}{54}}, \bibinfo{pages}{17954} (\bibinfo{year}{1996}).

\bibitem[{\citenamefont{Brey and Fertig}(2006)}]{Brey_2006b}
\bibinfo{author}{\bibfnamefont{L.}~\bibnamefont{Brey}} \bibnamefont{and}
  \bibinfo{author}{\bibfnamefont{H.}~\bibnamefont{Fertig}},
  \bibinfo{journal}{Phys.\ Rev.\ B} \textbf{\bibinfo{volume}{73}},
  \bibinfo{pages}{195408} (\bibinfo{year}{2006}).

\bibitem[{\citenamefont{Chico et~al.}(1996)\citenamefont{Chico, Benedict,
  Louie, and Cohen}}]{Chico_1996b}
\bibinfo{author}{\bibfnamefont{L.}~\bibnamefont{Chico}},
  \bibinfo{author}{\bibfnamefont{L.~X.} \bibnamefont{Benedict}},
  \bibinfo{author}{\bibfnamefont{S.~G.} \bibnamefont{Louie}}, \bibnamefont{and}
  \bibinfo{author}{\bibfnamefont{M.~L.} \bibnamefont{Cohen}},
  \bibinfo{journal}{Phys.\ Rev.\ B} \textbf{\bibinfo{volume}{54}},
  \bibinfo{pages}{2600} (\bibinfo{year}{1996}).

\bibitem[{\citenamefont{Chico et~al.}(1998)\citenamefont{Chico, L\'opez-Sancho,
  and Mu\~noz}}]{Chico_1998}
\bibinfo{author}{\bibfnamefont{L.}~\bibnamefont{Chico}},
  \bibinfo{author}{\bibfnamefont{M.~P.} \bibnamefont{L\'opez-Sancho}},
  \bibnamefont{and} \bibinfo{author}{\bibfnamefont{M.}~\bibnamefont{Mu\~noz}},
  \bibinfo{journal}{Phys. Rev. Lett.} \textbf{\bibinfo{volume}{81}},
  \bibinfo{pages}{1278} (\bibinfo{year}{1998}).

\bibitem[{\citenamefont{Santos et~al.}(2009)\citenamefont{Santos, Chico, and
  Brey}}]{Santos_2009}
\bibinfo{author}{\bibfnamefont{H.}~\bibnamefont{Santos}},
  \bibinfo{author}{\bibfnamefont{L.}~\bibnamefont{Chico}}, \bibnamefont{and}
  \bibinfo{author}{\bibfnamefont{L.}~\bibnamefont{Brey}},
  \bibinfo{journal}{Phys. Rev. Lett.} \textbf{\bibinfo{volume}{103}},
  \bibinfo{pages}{086801} (\bibinfo{year}{2009}).

\bibitem[{\citenamefont{Nilsson et~al.}(2007)\citenamefont{Nilsson,
  Castro~Neto, Guinea, and Peres}}]{Nilsson_2007}
\bibinfo{author}{\bibfnamefont{J.}~\bibnamefont{Nilsson}},
  \bibinfo{author}{\bibfnamefont{A.~H.} \bibnamefont{Castro~Neto}},
  \bibinfo{author}{\bibfnamefont{F.}~\bibnamefont{Guinea}}, \bibnamefont{and}
  \bibinfo{author}{\bibfnamefont{N.~M.~R.} \bibnamefont{Peres}},
  \bibinfo{journal}{Phys. Rev. B} \textbf{\bibinfo{volume}{76}},
  \bibinfo{pages}{165416} (\bibinfo{year}{2007}).

\bibitem[{\citenamefont{Castro et~al.}(2007)\citenamefont{Castro, Novoselov,
  Morozov, Peres, dos Santos, Nilsson, Guinea, Geim, and Neto}}]{Castro_2007}
\bibinfo{author}{\bibfnamefont{E.~V.} \bibnamefont{Castro}},
  \bibinfo{author}{\bibfnamefont{K.~S.} \bibnamefont{Novoselov}},
  \bibinfo{author}{\bibfnamefont{S.~V.} \bibnamefont{Morozov}},
  \bibinfo{author}{\bibfnamefont{N.~M.~R.} \bibnamefont{Peres}},
  \bibinfo{author}{\bibfnamefont{J.~M. B.~L.} \bibnamefont{dos Santos}},
  \bibinfo{author}{\bibfnamefont{J.}~\bibnamefont{Nilsson}},
  \bibinfo{author}{\bibfnamefont{F.}~\bibnamefont{Guinea}},
  \bibinfo{author}{\bibfnamefont{A.~K.} \bibnamefont{Geim}}, \bibnamefont{and}
  \bibinfo{author}{\bibfnamefont{A.~H.~C.} \bibnamefont{Neto}},
  \bibinfo{journal}{Phys. Rev. Lett.} \textbf{\bibinfo{volume}{99}},
  \bibinfo{pages}{216802} (\bibinfo{year}{2007}).

\bibitem[{\citenamefont{Snyman and Beenakker}(2007)}]{Snyman_2007}
\bibinfo{author}{\bibfnamefont{I.}~\bibnamefont{Snyman}} \bibnamefont{and}
  \bibinfo{author}{\bibfnamefont{C.~W.~J.} \bibnamefont{Beenakker}},
  \bibinfo{journal}{Phys. Rev. B} \textbf{\bibinfo{volume}{75}},
  \bibinfo{pages}{045322} (\bibinfo{year}{2007}).

\bibitem[{\citenamefont{Oostinga et~al.}(2007)\citenamefont{Oostinga, Heersche,
  Liu, Morpurgo, and Vandersypen}}]{Oostinga_2007}
\bibinfo{author}{\bibfnamefont{J.~B.} \bibnamefont{Oostinga}},
  \bibinfo{author}{\bibfnamefont{H.~B.} \bibnamefont{Heersche}},
  \bibinfo{author}{\bibfnamefont{X.}~\bibnamefont{Liu}},
  \bibinfo{author}{\bibfnamefont{A.~F.} \bibnamefont{Morpurgo}},
  \bibnamefont{and} \bibinfo{author}{\bibfnamefont{L.~M.~K.}
  \bibnamefont{Vandersypen}}, \bibinfo{journal}{Nature Materials}
  \textbf{\bibinfo{volume}{7}}, \bibinfo{pages}{151} (\bibinfo{year}{2007}).

\bibitem[{\citenamefont{Fiori and Iannaccone}(2009)}]{Fiori_2009a}
\bibinfo{author}{\bibfnamefont{G.}~\bibnamefont{Fiori}} \bibnamefont{and}
  \bibinfo{author}{\bibfnamefont{G.}~\bibnamefont{Iannaccone}},
  \bibinfo{journal}{Electron Device Letters, IEEE}
  \textbf{\bibinfo{volume}{30}}, \bibinfo{pages}{261 } (\bibinfo{year}{2009}).

\bibitem[{\citenamefont{Russo et~al.}(2009)\citenamefont{Russo, Craciun,
  Yamamoto, Tarucha, and Morpurgo}}]{Russo_2009}
\bibinfo{author}{\bibfnamefont{S.}~\bibnamefont{Russo}},
  \bibinfo{author}{\bibfnamefont{M.}~\bibnamefont{Craciun}},
  \bibinfo{author}{\bibfnamefont{M.}~\bibnamefont{Yamamoto}},
  \bibinfo{author}{\bibfnamefont{S.}~\bibnamefont{Tarucha}}, \bibnamefont{and}
  \bibinfo{author}{\bibfnamefont{A.}~\bibnamefont{Morpurgo}},
  \bibinfo{journal}{New J.Phys.} \textbf{\bibinfo{volume}{11}},
  \bibinfo{pages}{095018} (\bibinfo{year}{2009}).

\bibitem[{\citenamefont{Gonz\'alez et~al.}(2010)\citenamefont{Gonz\'alez,
  Santos, Pacheco, Chico, and Brey}}]{Gonzalez_2010}
\bibinfo{author}{\bibfnamefont{J.~W.} \bibnamefont{Gonz\'alez}},
  \bibinfo{author}{\bibfnamefont{H.}~\bibnamefont{Santos}},
  \bibinfo{author}{\bibfnamefont{M.}~\bibnamefont{Pacheco}},
  \bibinfo{author}{\bibfnamefont{L.}~\bibnamefont{Chico}}, \bibnamefont{and}
  \bibinfo{author}{\bibfnamefont{L.}~\bibnamefont{Brey}},
  \bibinfo{journal}{Phys. Rev. B} \textbf{\bibinfo{volume}{81}},
  \bibinfo{pages}{195406} (\bibinfo{year}{2010}).

\bibitem[{\citenamefont{McCann and Fal\char39{}ko}(2006)}]{McCann_2006}
\bibinfo{author}{\bibfnamefont{E.}~\bibnamefont{McCann}} \bibnamefont{and}
  \bibinfo{author}{\bibfnamefont{V.~I.} \bibnamefont{Fal\char39{}ko}},
  \bibinfo{journal}{Phys. Rev. Lett.} \textbf{\bibinfo{volume}{96}},
  \bibinfo{pages}{086805} (\bibinfo{year}{2006}).

\bibitem[{\citenamefont{Castro et~al.}(2008)\citenamefont{Castro, Peres,
  Lopes~dos Santos, Neto, and Guinea}}]{Castro_2008}
\bibinfo{author}{\bibfnamefont{E.~V.} \bibnamefont{Castro}},
  \bibinfo{author}{\bibfnamefont{N.~M.~R.} \bibnamefont{Peres}},
  \bibinfo{author}{\bibfnamefont{J.~M.~B.} \bibnamefont{Lopes~dos Santos}},
  \bibinfo{author}{\bibfnamefont{A.~H.~C.} \bibnamefont{Neto}},
  \bibnamefont{and} \bibinfo{author}{\bibfnamefont{F.}~\bibnamefont{Guinea}},
  \bibinfo{journal}{Phys. Rev. Lett.} \textbf{\bibinfo{volume}{100}},
  \bibinfo{pages}{026802} (\bibinfo{year}{2008}).

\bibitem[{\citenamefont{Martin et~al.}(2008)\citenamefont{Martin, Blanter, and
  Morpurgo}}]{Martin_2008}
\bibinfo{author}{\bibfnamefont{I.}~\bibnamefont{Martin}},
  \bibinfo{author}{\bibfnamefont{Y.~M.} \bibnamefont{Blanter}},
  \bibnamefont{and} \bibinfo{author}{\bibfnamefont{A.~F.}
  \bibnamefont{Morpurgo}}, \bibinfo{journal}{Phys. Rev. Lett.}
  \textbf{\bibinfo{volume}{100}}, \bibinfo{pages}{036804}
  (\bibinfo{year}{2008}).

\bibitem[{\citenamefont{A.Morpurgo}()}]{Morpurgo}
\bibinfo{author}{\bibnamefont{A.Morpurgo}}, \bibinfo{note}{talk given at the
  Graphene Week 2010. April 2010, College Park, MD.}

\bibitem[{\citenamefont{Datta}(1995)}]{Datta_book}
\bibinfo{author}{\bibfnamefont{S.}~\bibnamefont{Datta}},
  \emph{\bibinfo{title}{Electronic Transport in Mesoscopic Systems}}
  (\bibinfo{publisher}{Cambridge University Press},
  \bibinfo{address}{Cambridge}, \bibinfo{year}{1995}).

\bibitem[{\citenamefont{Gonz\'alez et~al.}(2009)\citenamefont{Gonz\'alez,
  Rosales, and Pacheco}}]{Gonzalez_2009}
\bibinfo{author}{\bibfnamefont{J.}~\bibnamefont{Gonz\'alez}},
  \bibinfo{author}{\bibfnamefont{L.}~\bibnamefont{Rosales}}, \bibnamefont{and}
  \bibinfo{author}{\bibfnamefont{M.}~\bibnamefont{Pacheco}},
  \bibinfo{journal}{Physica B: Condensed Matter}
  \textbf{\bibinfo{volume}{404}}, \bibinfo{pages}{2773 }
  (\bibinfo{year}{2009}).

\bibitem[{\citenamefont{Chico and Jask\'{o}lski}(2004)}]{Chico_2004}
\bibinfo{author}{\bibfnamefont{L.}~\bibnamefont{Chico}} \bibnamefont{and}
  \bibinfo{author}{\bibfnamefont{W.}~\bibnamefont{Jask\'{o}lski}},
  \bibinfo{journal}{Phys. Rev. B} \textbf{\bibinfo{volume}{69}},
  \bibinfo{eid}{085406} (\bibinfo{year}{2004}).

\end{thebibliography}
\end{document}